\documentstyle[psfig]{mn}

\ifnfsstwo

\fi

\ifnfssone
  \newmathalphabet{\mathit}
    \addtoversion{normal}{\mathit}{cmr}{m}{it}
    \addtoversion{bold}{\mathit}{cmr}{bx}{it}

\fi

\ifoldfss

\fi

\newcommand{\micron}{$\mu$m}

  \makeatletter
  \ifx\CUP@mtlplain@loaded\undefined  
  \makeatother  
    
  \else
    
  \fi

\loadboldmathitalic 
\loadboldgreek      
  
\makeatletter
\ifx\CUP@mtlplain@loaded\undefined  
  \newfont\bit{cmbxti10 at 9pt}
\else
  \newfont\bit{mtbxti10 at 9pt}
\fi
\makeatother  
 
\title[IRAS 4] {NGC1333/IRAS4: A multiple star formation laboratory} \author[Smith et al.]
{K.W. Smith$^{1,2,3}$, I.A. Bonnell$^{4,5}$, J.P. Emerson$^3$ \& T. Jenness$^{6}$ 
\\ 
$^{1}$ Institut f\"ur Astronomie, ETH-Zentrum, CH-8092 Z\"urich, Switzerland\\
$^{2}$ Paul Scherrer Institut, W\"urenlingen und Villigen, CH-5232 
Villigen PSI, Switzerland\\
$^{3}$ Physics Department, Queen Mary \&Westfield College, Mile End Road, London E1 4NS, UK. \\ 
$^{4}$ Institute of Astronomy, Madingley Road,
Cambridge CB3 0HA, UK. \\
$^{5}$ University of St Andrews, School of Physics and Astronomy,  North Haugh, 
St Andrews, Fife, KY16 9SS, Scotland \\
$^{6}$ Joint Astronomy Centre, 660 N.\ A`oh\={o}k\={u} Pl., 
Hilo, HI 96720, USA }
\date{Received 1 October 1999, Accepted }
\pagerange{\pageref{firstpage}--\pageref{lastpage}}
\pubyear{    }

\def\LaTeX{L\kern-.36em\raise.3ex\hbox{a}\kern-.15em
    T\kern-.1667em\lower.7ex\hbox{E}\kern-.125emX}

\begin{document}

\label{firstpage}

\maketitle

\begin{abstract}

We present SCUBA observations of the protomultiple system NGC1333/IRAS4 at
450\micron\ and 850\micron.  The 850\micron\ map shows significant
extended emission which is most probably a remnant of the initial cloud
core. At 450\micron, the component 4A is seen to have an elongated shape
suggestive of a disk. Also we confirm that in addition to the 4A and 4B
system, there exists another component 4C, which appears to lie out of the
plane of the system and of the extended emission.  Deconvolution of the
beam reveals a binary companion to IRAS4B.  Simple considerations of
binary dynamics suggest that this triple 4A-4BI-4BII system is unstable
and will probably not survive in its current form. Thus IRAS4 provides
evidence that systems can evolve from higher to lower multiplicity as they
move towards the main sequence.  We construct a map of spectral index from
the two wavelengths, and comment on the implications of this for dust
evolution and temperature differences across the map.  There is evidence
that in the region of component 4A the dust has evolved, probably by
coagulating into larger or more complex grains. Furthermore, there is
evidence from the spectral index maps that dust from this object is being
entrained in its associated outflow.

\end{abstract}


\section{Introduction}

Stars of all ages are commonly found in binary and multiple systems
\cite{duqmay}. In fact, amongst the youngest stars, the frequency of
companions appears to be higher than in older systems (eg Ghez, 1995). 
In order to understand the initial stages of star
formation, we need to understand the initial stages of binary star
formation. There have been many theories advanced to explain the
formation of binary stars (cf Clarke 1995; Bonnell
1999). These generally include either a fragmentation
during collapse \cite{boss,bonnetal}, a fragmentation of a
circumstellar disc \cite{bonnfrag,whitworth} or a post-fragmentation
star-disc capture (eg Clarke \& Pringle 1991). In
order to be able to distinguish between these theories, we need to
observe the youngest systems. This paper reports on recent
observations of a young, protostellar multiple system NGC~1333 IRAS~4 in order
to constrain its formation mechanism.

NGC1333/IRAS4 is a well studied protobinary system.
It was first identified as a site of star formation 
by Haschick et al (1986), who
observed two variable H$_2$O masers. The distinct core was mapped
by Jennings et al (1987). 

Sandell et al (1991, hereafter SADRR) mapped the system in the
submillimetre with UKT14. They confirmed its multiple nature, finding
a 30'' binary system embedded in diffuse emission. They labelled the
components 4A and 4B. Component 4A was seen to be elongated, which
SADRR interpreted as a massive disk seen at an oblique angle.
Dynamical evidence of binarity for IRAS4 is lacking, but the common
envelope of material surrounding the two main components suggests that
they are the products of a single collapse event, the extended
material being identified as the remnants of the precollapse
core. This extended envelope then provides the main motivation for
considering IRAS 4 to be a multiple system, rather than a chance
superposition of cores.

The system is associated with a high velocity outflow, mapped at
various molecular transitions by Blake et al (1995, hereafter BSDGMA).  These
authors found a high velocity outflow originating from IRAS 4A and
aligned with the apparent disk axis.

Minchin et al \shortcite{minchin} measured polarisation at 800\micron\ 
for both 4A and 4B. They found that the polarisation position angle
was similar for both sources and broadly aligned with the
elongated circumbinary emission.

Interferometry by Lay et al \shortcite{lay} revealed that 
both 4A and 4B are themselves multiple systems. 4A was revealed to
be a binary of separation 1.2'' aligned with the direction of elongation
of 4A. 4B appeared also to be multiple, but had a more complex
nature which could not be determined.

\section{Observations}

The observations were taken with SCUBA (Submillimetre Common-User
Bolometer Array) at the James Clerk Maxwell Telescope on 1997 July 6 about
2 hours after sun rise. "Jiggle map" data were obtained simultaneously at
two wavelengths with SCUBA (Holland et al. 1999) using two hexagonal
arrays of 91 bolometers at 450\micron\ and 37 bolometers at 850\micron\
with a field of view of about 2.3 arcmin. The total integration time
(on+off) was approximately 60 minutes (20 integrations).

The SURF (SCUBA User Reduction Facility; Jenness \& Lightfoot 1998)
package was used for flat-fielding, extinction correction, sky noise
removal, despiking, removal of bad pixels, rebinning and calibration
of the images. The sky opacity at zenith was calculated to be 0.38 at
850\micron\ from skydip observations and 2.1 at 450\micron\ using the
standard extrapolation formula (Jenness et al, in prep). Saturn was
used for calibration at 850\micron, resulting in a flux 
conversion factor of 285 Jy/beam/V, but was too large to be used for
450\micron\ (no other planet was available), therefore a flux
conversion factor of 1200 Jy/beam/V was inferred for the 450\micron\
observations by examining calibration observations taken at
approximately the same time of day from other nights.

\section{Morphology of IRAS4 and comparison with earlier results}

\begin{figure}
\vspace{0.25in}
\hspace{0.3in}\psfig{{figure=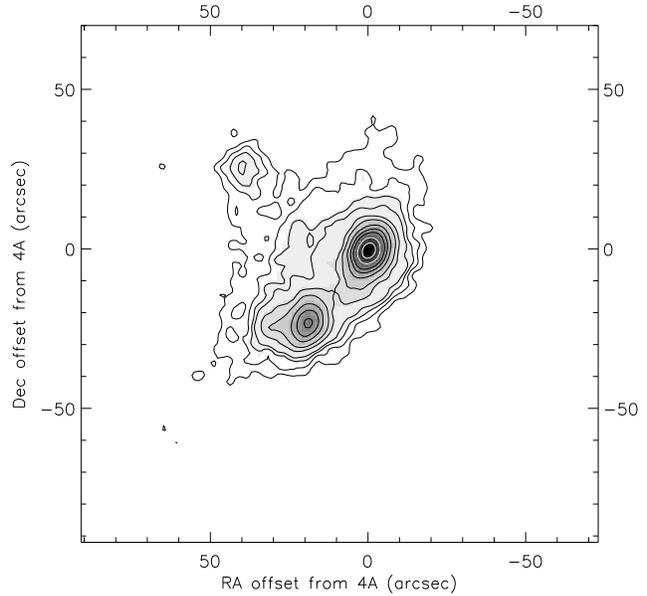,width=2.7truein,height=2.7truein}}
\vspace{0.25in}
\caption{\label{450cont} Contour plot of 450\micron\ jiggle map,
overlaid on a greyscale image. The noise is estimated to be
approximately 1 Jy/beam, from the RMS deviation about the mean
measured in an off source region (marked in Figure~\ref{regions}. The
FWHM of the beam is approximately 9''.  Contours are drawn at 3, 4, 5,
6, 9, 12, 18, 24, 30, 40, 50 and 60 times the noise level of
0.9Jy/beam. Contours near the peaks have been plotted in white so they can
be distinguished from the background.
Offsets in RA and Dec are in arcseconds from the peak of
4A, at $\alpha$(2000)=3$^h$ 29$^m$ 10.4$^s$, 
$\delta$(2000)=31$^o$~13$^{'}$~33.6$^{''}$ }
\end{figure}

\begin{figure}
\vspace{0.25in}
\hspace{0.3in}\psfig{{figure=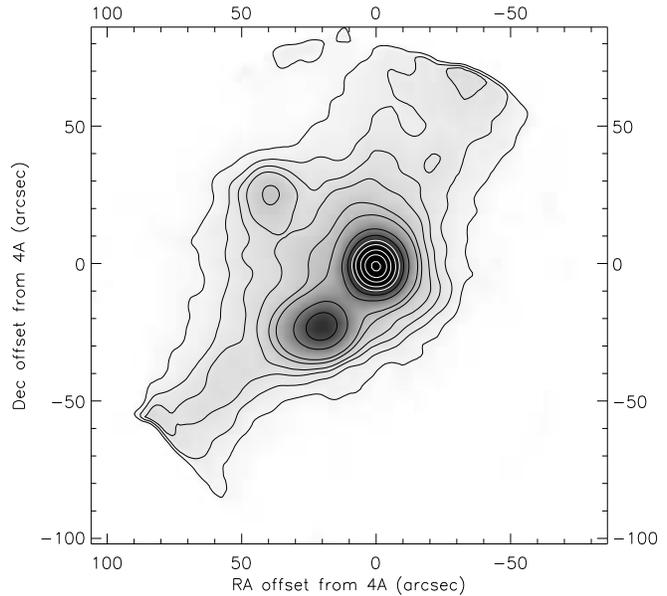,width=2.7truein,height=2.7truein}}
\vspace{0.25in}
\caption{\label{850cont} Contour plot and greyscale of 850\micron\
jiggle map. The noise is estimated to be 0.05 Jy/beam.  The FWHM of
the beam is approximately 16''. Contours are drawn at 3, 6, 9, 12, 18,
24, 48, 72, 96, 120, 150, 175 and 200 times the noise level. 
Offsets are
measured from the peak of 4A at $\alpha$(2000)=3$^h$ 29$^m$ 10.3$^s$, 
$\delta$(2000)=31$^o$ 13$^{'}$ 32.9$^{''}$ }
\end{figure}

Both jiggle maps are shown as contour plots in Figures~\ref{450cont}
and~\ref{850cont}. The two distinct components of IRAS4 are clear in
both Figures.  The main components lie along an axis running
NW-SE. The position angle of this axis is approximately 130$^o$ (where
position angle is taken to be zero for objects oriented North-South
and increases anticlockwise) The position angle of the extended
emission surrounding the main system is difficult to determine
accurately from the jiggle maps, since the emission region is slightly
larger than the size of the map in the long axis direction, and edge
effects come into play. However, the direction of elongation is
broadly the same as the position angle of the 4A-4B axis.  The third
knot of emission is visible in both maps, lying some 49 arcseconds to
the NE of the 4A-4B axis. This was also noted by SADRR. We will
henceforth refer to this as '4C'.

\begin{figure}
\psfig{{figure=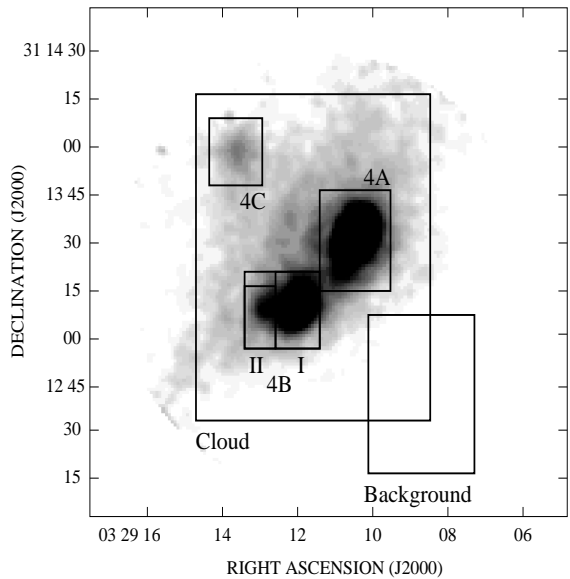,width=3.truein,height=3.truein}}
\caption{\label{regions} Inverted greyscale plot of raw 450\micron\
map, with the regions used for photometry for 4A, 4BI, 4BII, and 4C
marked. The region used at 850\micron\ for all 4B is also marked. The
region at lower right was used to calculate the background level. The
large region covering the whole central part is that used to calculate
the extended emission, as discussed in Section~\ref{extended}. }
\end{figure}

The peak flux densities we measure from our maps are listed in
Table~\ref{rawflux}, together with values from SADRR for comparison.  
The errors of SADRR are statisitical errors, taking no account of
calibration uncertainties. The errors we quote are RMS deviations from the 
mean measured in an off-source region (see Figure~\ref{regions}). 
There will also be considerable calibration 
uncertainty in both sets of measurements.

At 850\micron\ and 800\micron\ the fluxes agree well for source 4A,
but not so well for 4B. Since our 850\micron\ calibration was
calibrated reliably, and it is unlikely that such a large discrepency
is due to the small difference in the waveband used, we attribute this
discrepency to the UKT14 measurements being made one point at a time,
with sky conditions and other factors changing during the
observations.

At 450\micron\ the fluxes agree well for source 4B but not for 4A.  At
this wavelength, calibration inaccuracies for our fluxes could be
large. The agreement in flux for source 4B is encouraging, suggesting
that our calibration error is likely to be relatively small. Even the
4A flux is not discrepent by as much as a factor of 2. Averaging the
discrepencies between the fluxes of these two objects suggests that
our calibration error could be around 30\%. We note again that the
difference in observing technique between UKT14 and SCUBA means that
our data is likely to be more internally self-consistent than the
UKT14 data since the whole field is observed simultaneously with
SCUBA.

\begin{table*}
\begin{center}
\begin{tabular}{c|c|c} \hline
\multicolumn{3}{l}{450\micron} \\
Object & Peak flux density (Sandell et al.) & Peak flux density (current study) \\
       &     (Jy / beam)                    &  Jy / beam          \\
4A     &   35.6 $\pm$ 4.2                    & 51.03 $\pm$ 1.0     \\
4B     &   28.0 $\pm$ 5.0                    & 28.04 $\pm$ 1.0    \\        
4C     &   -                                 & 5.56 $\pm$  1.0 \\ \hline
\multicolumn{3}{l}{850\micron / 800\micron } \\
4A     &  10.9 $\pm$ 0.3                     & 10.3 $\pm$ 0.03  \\
4B     &   5.76 $\pm$ 0.15                   & 4.5  $\pm$ 0.03   \\
4C     &   -                                 & 0.97 $\pm$ 0.03  \\
\end{tabular}
\caption{\label{rawflux} The peak flux densities from our maps, and
those found by Sandell et al. (1991). These authors observed at
450\micron\ and 800\micron\ using UKT14, with beam sizes of 13.5'' HPBW
(800\micron) and 7.7'' HPBW (450\micron). The chop throw used by
Sandell was 60'' in both cases. The quoted errors on our values
represent the RMS deviation from the mean measured in an off-source
region of our maps (marked on Figure~\ref{regions}).  }
\end{center}
\end{table*}

From the 450\micron\ plot in Figure~\ref{450cont} we can see that 4A
itself is significantly elongated, with long axis at a position angle
of 154$^o$, slightly larger than the position axis of 4A-4B.
Component 4B is also seen to be elongated, this time in an EW
direction. 

Interferometric studies (Lay et al, 1995) have revealed that 4A is
itself a binary, with separation of 1.2'' and position angle aligned
with the elongation.  The natural interpretation of the large
elongated 4A system is that it is a circumbinary disk, foreshortened
in the plane of the sky, although from the continuum maps it is
possible that it could be an elongated spheroid.  We can fit an
elliptical Gaussian to 4A and derive the likely inclination angle of
the disk. The best fit elliptical Gaussian has position angle 154$^o$,
and ellipticity 0.26, where the ellipticity is defined as $(1-b/a)$,
with $a$ and $b$ being the major and semimajor axes respectively. From
this we can infer a likely inclination to the plane of the sky of
42$^o$. The elongation of 4A was also seen by SADRR. They too
inferred the presence of a tilted disk and derived a similar apparent
inclination to the plane of the sky. The molecular line observations
of BSDGMA show bipolar outflows from 4A, whose alignment on the
sky and kinematic properties point to them being emitted roughly
perpendicular to the plane of this disk.  The question of the 4A disk
will be addressed again in Section~\ref{lucy}.

Lay et al's interferometry implied that 4B is a multiple system, but
they could not determine the exact nature of 4B as the system appeared
to be too complex. In both our submillimetre maps, 4B appears extended
in an E-W direction. In the 450\micron\ map this extension is in the
form of a small lobe, jutting out from the main part of 4B.  This
appearance suggests that we may have almost resolved a second
component of the 4B system. We will discuss this issue further in
Section~\ref{lucy}.

\section{Richardson-Lucy deconvolution.}
\label{lucy}

The high signal to noise of our maps, together with the existence of
high signal to noise planetary maps taken during the same night,
allows us to deconvolve the measured SCUBA beam from the data. We did
not use our Saturn maps for this, as it is not clear that Saturn is
circular in the submillimetre. We instead used high signal-to-noise
maps of Uranus taken several hours earlier. We chose to use a
Richardson-Lucy technique (Lucy, 1974), as this conserves the flux
density. We used a version implemented in the {\tt IRAF} {\em stsdas}
package\footnote{{\tt IRAF} is distributed by the National Optical
Astronomy Observatories, which are operated by AURA under co-operative
agreement with the National Science Foundation.}.

\subsection{The beam maps}
\label{beams}

Inspection of the reduced beam maps revealed a departure
from circular symmetry at the level of a few percent, with the
non-circular part of the beam roughly aligned with the direction of
the chop throw (i.e. azimuth). This indicates that the non-circularity due
to the chopping dominated over any non-circularity inherent in the array.
The beam maps were therefore rotated so that they were azimuthally-aligned 
with the IRAS4 maps.

\begin{figure}
\hbox{
\psfig{{figure=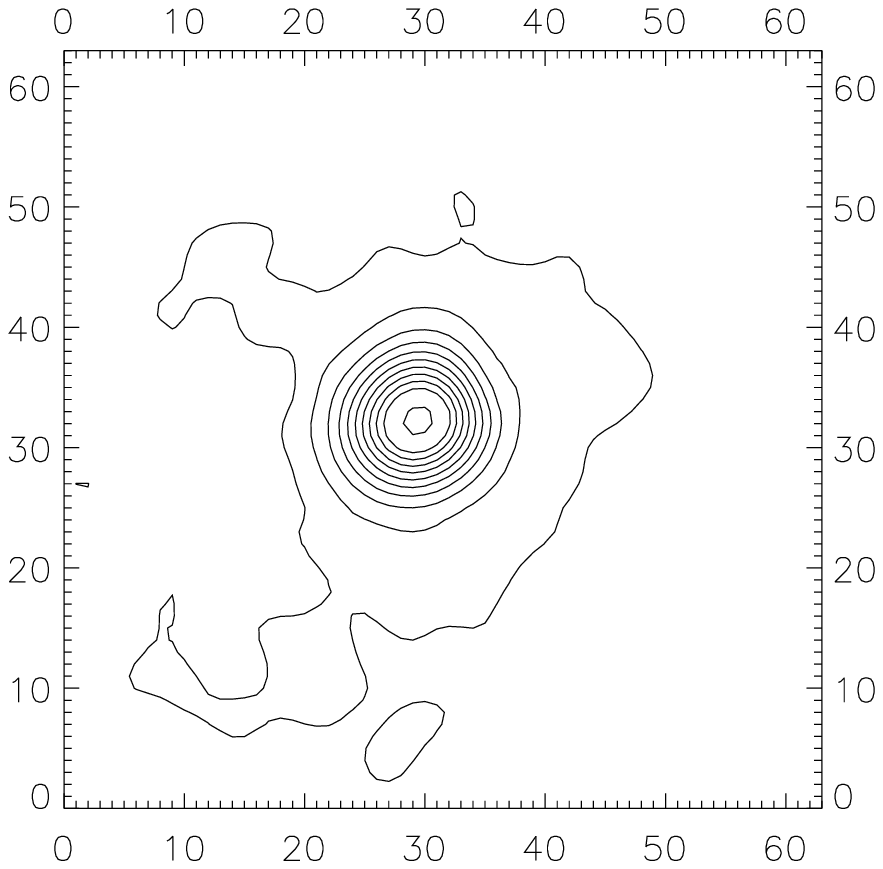,width=1.5truein,height=1.5truein}}
\psfig{{figure=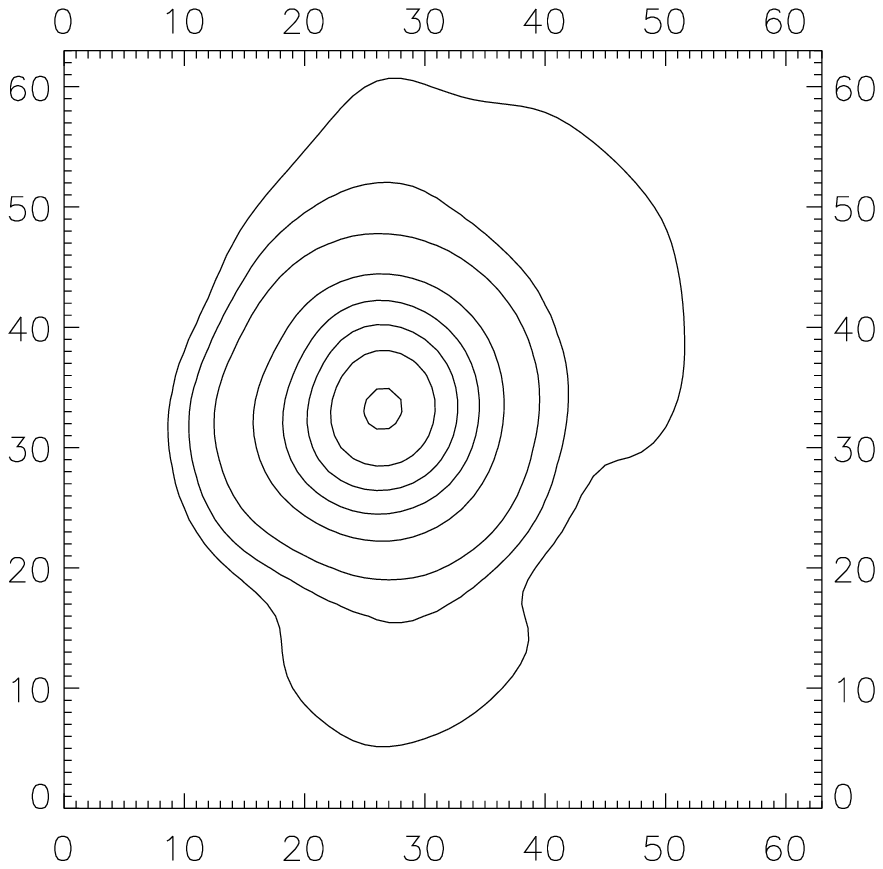,width=1.5truein,height=1.5truein}}}
\caption{\label{beamspic} Contour plots of beam maps, 450\micron\
(left) and 850\micron\ (right). The lowest contour at 450\micron\ is
at 10\% of the peak flux, the lowest contour at 850\micron\ is at 2\%
of the peak flux. These contours show the level of the non-circular
part of the beam power. Both beam maps have been rotated to the same
azimuthal orientation as the object maps. X and Y axes are in
arcseconds.}
\end{figure}

The beam maps were apodized by deconvolving a smoothed, circular
image, with the radius of Uranus. This removes the planet image from
the map, leaving us with just the beam. Both beam maps are shown in
Figure~\ref{beamspic}.

We measured the effective beam area from the Uranus beam maps. The
effective area of the 450\micron\ beam was found to be 110 arcsec$^2$,
and that of the 850\micron\ beam was 290 arcsec$^2$. The half-power
beam widths, measured by fitting circular Gaussians to the beams, were
8.9'' and 15.8'' for 450\micron\ and 850\micron\ respectively. The
850\micron\ HPBW is therefore consistent with the measured beam area to
within a few percent.  The 450\micron\ fitted HPBW implies a beam area
of 90 arcsec$^2$, 20\% smaller than the measured beam area. This gives
us an estimate of the effective area of the extended 450\micron\ error beam.

\subsection{Stopping criteria}
\label{crits}

The decision of when to terminate the deconvolution procedure is of
critical importance. Failure to follow the technique far enough
results in a non-fully deconvolved map. Following the process too far
will result in building noise into apparent structure. Noise levels
were measured in regions of the maps judged to best approximate
background sky emission. These noise levels could then be specified as
a constant noise applied to each pixel, and the deconvolution was
terminated when the $\chi^{2}$ calculated for the observed image and
the recovered image convolved with the beam reached 1.
This convergence criterion was found to be critically dependent on the
noise value in the case of the 450\micron\ map. Values of the RMS
deviation about the mean were measured in a region off the cloud
and found to range from 0.78 to 1.1 Jy/beam. 
Noise levels of around 1.0 were found 
to lead to rapid convergence and an image not
significantly improved from the original. Levels as low as 0.95 led to
convergence in many iterations, and a final image in which faint
structure around IRAS4C had built into peculiar artefacts in the
image. We therefore selected a noise level of 0.96. This led
to convergence in 48 iterations, and a final image which appeared
significantly enhanced compared to the original, but without any
apparently artificial structure.  It is important to stress that,
whilst the assumed noise level was entirely realistic for the
450\micron\ map, it was eventually selected because it leads to an
acceptable deconvolution.

\subsection{The deconvolved maps}

\begin{figure}
\vspace{0.25in}
\hspace{0.3in}\psfig{{figure=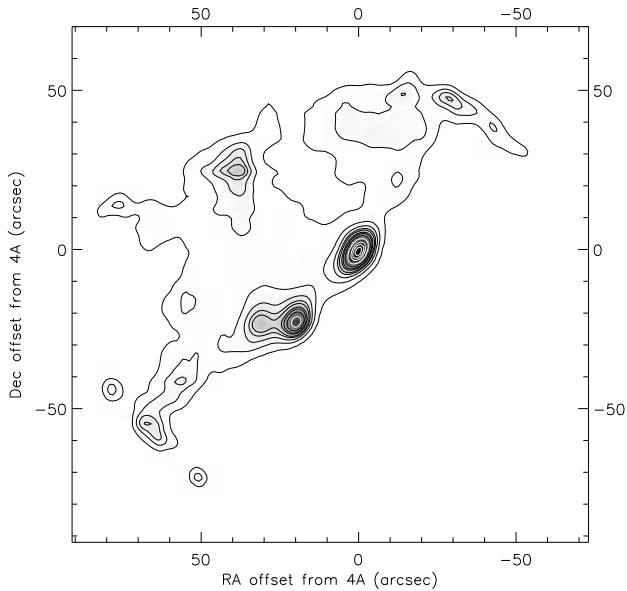,width=2.7truein,height=2.7truein}}
\vspace{0.25in}
\caption{\label{450decon} Contour plot of Richardson-Lucy deconvolved
450\micron\ map, overlaid on a greyscale image. Contours are drawn at
4.5$\times10^{-3}$, 4.5$\times10^{-2}$, 0.14, 0.28, 0.41, 0.68, 0.91, 1.4, 2.3,
3.6, 4.5 and 5.5 Jy arcsec$^{-2}$.}
\end{figure}

The results of the deconvolution can be seen in Figure~\ref{450decon}.
The most striking aspects of these images are the disappearance of
much of the circumbinary material, and the resolution of 4B
into a binary of separation 12''.

\begin{figure}
\vspace{0.25in}
\hspace{0.3in}\psfig{{figure=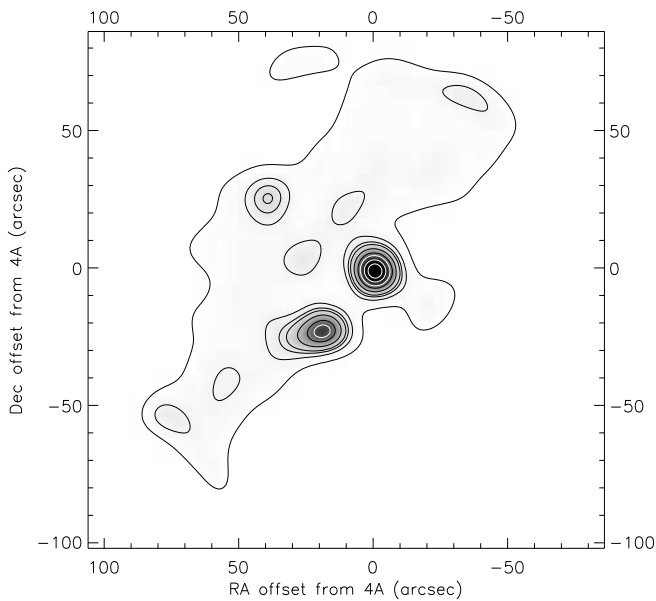,width=2.7truein,height=2.7truein}}
\vspace{0.25in}
\caption{\label{850decon} Contour plot
of Richardson-Lucy deconvolved 850\micron\ map. Contour levels are 
5$\times10^{-4}$, 2.6$\times10^{-3}$, 5$\times10^{-3}$, 7.8$\times10^{-3}$,
1.6$\times10^{-2}$, 0.03, 0.05, 0.06,
0.08 and 0.12 Jy arcsec$^{-2}$.}
\end{figure}

The 850\micron\ map noise level was estimated similarly and found to be
approximately 0.03 Jy/beam. The deconvolved 850\micron\ map is shown in
Figure~\ref{850decon}. The deconvolution was again halted when the
$\chi^{2}$ reached 1.

\begin{figure}
\psfig{{figure=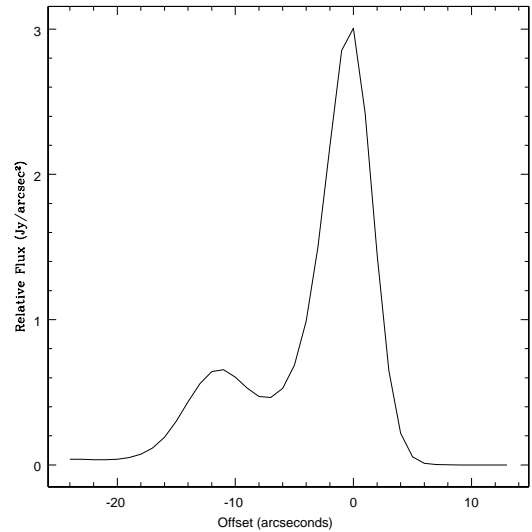,width=3.truein,height=3.truein}}
\caption{\label{4BI4BIIcut} A cut across
the deconvolved 450\micron\ map along the 4BI-4BII axis, connecting the
component peaks.  The offset in arcseconds is measured from the peak
position of component 4BI.}
\end{figure}

Gaussian fits were made to the deconvolved components.  The 450\micron\
image of 4A was found to be well fit by a Gaussian with ellipticity
0.40 and position angle 148.6$^o$. The greater ellipticity now suggests an
inclination of 53$^o$ to the plane of the sky. 4B is seen in the deconvolved maps to consist of two distinct components, which we label BI and BII. 
A cut across the 4BI-4BII system in the 450\micron\ map is shown in 
Figure~\ref{4BI4BIIcut}. The separate components
of 4B could each be fitted independently. 4BI was found to have an
ellipticity of 0.23, implying an inclination angle of 40$^o$ to the
plane of the sky.  The position angle of 4BI was 140.41$^o$, aligned
much more closely with the position angle of 4A and with the 4A-4B
axis, which in the deconvolved 450\micron\ map was found to be
135$^o$. This suggests that 4BI may also possess a disk. 4BII was found
to have no measurable ellipticity.  In each case, deconvolved and raw
maps, no evidence for elongation was seen in the case of 4C.

The 4BI-4BII binary is probably too wide (and too simple) to account
entirely for the findings of Lay et al (1995). Thus it is likely that
either 4BI or 4BII or both are themselves multiple.  

Gaussian fitting of the components in the 850\micron\ deconvolved map
revealed no evidence for elongation in the case of 4A. 4B was found to 
have ellipticity 0.23 with position angle 102$^o$, which is clearly due
to the 4BI - 4BII system being unresolved.

\section{Spectral index}
\label{specin}

With observations at two wavelengths, it is possible in principle to
compare the two flux densities and measure the submillimeter spectral
index, $\alpha$, point by point across the region of interest.  We
note first that, since we have assumed our 450\micron\ flux
density conversion factor, the absolute values we derive for the
spectral index, and any derived quantities, will be reliant on the
accuracy of this assumption. Relative values will still be valid,
however, and it is these which we are most interested in.

In practice, the measurement of spectral index is complicated by the
uncertain nature of the instrumental beam at 450\micron.  In general,
the instrumental profile consists of a central Gaussian-type beam, and
an extended error beam, which can account for up to half of the
effective beam area (See Section~\ref{beams}).  This will have the
systematic effect of reducing the spectral index of compact sources
relative to the surrounding cloud. Particular care must therefore be
taken when interpreting low values of $\alpha$ for the compact sources
as evidence of physical differences, such as temperature or dust
evolution effects.

One way to avoid these systematic effects would be to use the
deconvolved maps, since the beam pattern will in theory have been
removed from these. However, use of the deconvolved maps is not free
from systematic effects either. The problems with stopping criteria
discussed in Section~\ref{crits} are of particular importance, since
the extended emission is quite sensitive to exactly when the
deconvolution is terminated. 

We therefore believe the best way to calculate spectral indices for
the compact sources is to use the photometry presented in
Section~\ref{phot}, which was obtained by integrating over appropriate
regions in the deconvolved maps calibrated in Jy/arcsec$^2$.  Values
calculated in this way appear in Table~\ref{betas}, together with
values of $\beta$ calculated at various temperatures.  We have also
constructed spectral index maps made from the undeconvolved images, in
order to examine the extended structure.

\begin{figure}
\vspace{0.25in}
\hspace{0.3in}\psfig{{figure=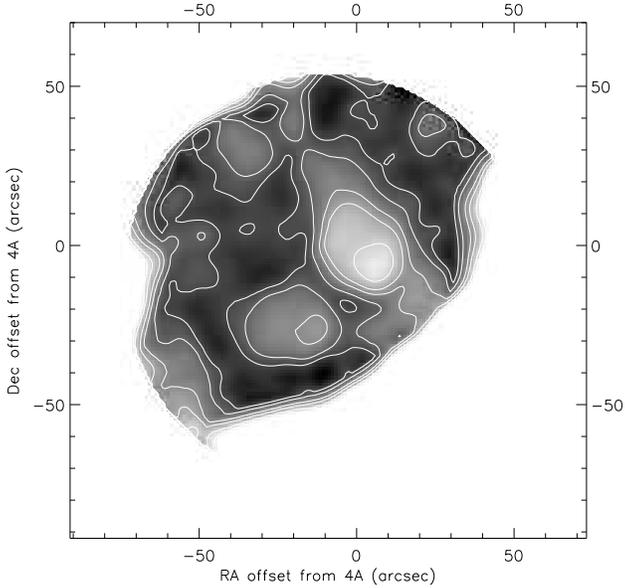,width=2.7truein,height=2.7truein}}
\vspace{0.25in}
\caption{\label{alphafig1} Spectral index map obtained from the raw
maps, shown as a greyscale plot with overlaid contours. Contour levels
are 3.0, 3.3, 3.5, 3.8, 4.0 and 4.5. The greyscale has been
inverted so that regions of higher spectral index appear darker.  The
x and y axes are arcsec offsets from the position of the peak of 4A.
Edge effects have been cut out.}
\end{figure}

Figure~\ref{alphafig1} shows the map of spectral index $\alpha$,
($F_\nu \propto \nu^{\alpha}$), computed from the raw maps.  This map
has been recalibrated into Jy/arcsec$^2$. The 450\micron\ map has
been degraded to the same resolution as 850\micron. This smoothing is
another cause of possible systematic effects, and the correct
smoothing was found by minimising the resulting artefacts by a process
of trial and error.

The RMS deviation from the mean of the cloud's spectral index,
measured in a region between 4A, 4B and 4C, is 0.08. This value was
used as an estimate of the random error in extended regions of the
spectral index map.

The compact sources are seen in Figure~\ref{alphafig1} to have lower
spectral index than the surrounding material. Values of $\alpha$ for
the sources in the SI map are typically 3 for 4A or 3.5 for 4B and 4C,
compared to 4.1 for the cloud.

A region of low spectral index is seen extending from 4A, roughly
perpendicular to the 4A disk axis. This region corresponds to the
outflow seen in molecular line emission by BSDGMA. The difference in
$\alpha$ between the outflow region and the surrounding cloud is
approximately 0.5, and is therefore significant compared to the RMS
error for the cloud. Since the outflow is not a compact source, it is
difficult to see how beam effects could have produced this result. We
conclude that the 4A outflow has a lower spectral index than the
surrounding cloud.

Computing values for $\alpha$ from the photometry of
Section~\ref{phot}, and comparing these to values measured for the
cloud from the SI map discussed above, we see again
that 4A has a significantly lower $\alpha$ than the cloud. The 4B
$\alpha$ is comparable to the cloud, and 4C is even a little
higher. This gives us more confidence that the 4A spectral index is
genuinely low, but lends no support to the idea that 4B or 4C have
significantly lower spectral index than the cloud.

Below, we discuss the various physical causes which 
could give rise to these spectral index variations.

\begin{table*}
\begin{center}
\begin{tabular}{c|c|c|c|c|c|c} \hline
Object      & $\alpha$ & $\beta$, (T=20K) & $\beta$, (T=30K) & $\beta$, (T=50K) & $\beta$, (T=100K) \\
Cloud       &  4.1$^*$&  2.8            &  2.5              &   {\bf 2.3}      &  {\bf 2.2 }   \\
4A          &  3.7    &  2.4            &  {\bf 2.1 }       &   1.9           &  1.8 \\
4A SW-NE    &  3.6$^*$&  2.3            &  2.0              &   {\bf 1.8}     &  {\bf 1.7}  \\
4B (I + II) &  3.9    &  2.6            &  {\bf 2.3 }       &   2.1           &  2.0 \\
4C          &  4.4    &  3.1            &  {\bf 2.9 }       &   2.7           &  2.5 \\ \hline
\end{tabular}
\caption{\label{betas} Values of spectral index, $\alpha$, for
different objects and regions in the raw spectral index map.  Values
for the compact sources are computed from the aperture photometry of
the deconvolved maps presented in Section~\ref{phot}, except those
marked $^*$ which are mean values found from regions of the raw
maps. In the case of the cloud, this region is a square lying near the
centre of the field between the three main sources and the outflow.
Values of $\beta$ are derived based on the assumption that the
emission is optically thin, using various values of the
temperature. Values subsequently used to calculate masses in
Tables~\ref{peakmass} and~\ref{extmass} are highlighted in bold.}
\end{center}
\end{table*}


\subsection*{Temperature}

Provided we assume the emission is optically thin, we can write the
spectral index as $\alpha = 2 + \beta + \gamma$, where $\beta$ is the
dust opacity spectral index and $\gamma$ is a factor taking account of
the departure from the Rayleigh-Jeans law at low temperatures, which
is always negative and becomes significant for temperatures below
about 50K. We have taken account of this correction when calculating
the $\beta$ values presented in the various tables and used in
Section~\ref{phot}.


The temperature of the IRAS4 sources was estimated by SADRR to be 33K,
based on a fit to the SED measured at millimetre and submillimetre
wavelengths. A temperature of around 30K therefore seems most likely
for the compact sources. 
The gas temperatures were estimated by BSDGMA from their
molecular line data, and lie in the range 20-40K for the compact
sources, and 70-100K for the wings (the outflow). The surrounding
cloud temperature is not estimated, but is likely to lie in between
these two values. 
To illustrate the effects
of different temperatures, we have calculated $\beta$ 
for several different cases, and we have 
used temperatures of 50K and 100K when calculating the cloud mass.

The difference in $\alpha$ between 4A and the cloud is 0.4.  To
account for this solely by a difference in temperature, the cloud
would need a temperature of approximately 100K if 4A has a temperature
of 30K, or 50K if 4A has a temperature of 20K. Thus it is possible to
explain the difference in spectral index for 4A as a temperature
effect, but this requires a lower temperature for 4A than SADRR
determined, or a cloud temperature similar to the temperature in the
outflow. The difference in $\alpha$ between the outflow and the cloud
is unlikely to be due to temperature, since the outflow is unlikely to be
significantly cooler than the cloud.

\subsection*{Optical depth}

Determination of spectral indices, and also inferences about the system's 
mass, depend on the emission being optically thin. We can 
investigate the validity of this assumption to some extent by calculating 
lower limits for the optical depth at each source. This is done by 
comparing the flux density observed to that expected from a blackbody 
at an assumed temperature (see e.g. Visser et al, 1998).  
For reasonable temperatures in the range 20-100K at
450\micron\ the optical depth at the position of the peak of 4A is at
least 0.09. At 850\micron, the peak of 4A has $\tau_{\nu} \geq
0.02$. There is therefore no evidence that our assumption of low
optical depth is invalid.

\subsection*{Line contamination}
\label{lines}

The principal components, and the outflow, are molecular line emission
sources (BSDGMA, Lefloch et al, 1998).  Contamination of the
850\micron\ band by line emission, particularly CO 3-2 at 870\micron,
could lead to lower spectral index in these regions. We can
investigate this possibility by converting BSDGMA's integrated
brightness temperatures for the sources into flux densities.  The
integrated brightness temperature of the CO 3-2 line is given as 135.4
K km s$^{-1}$ for 4A and 64.2 K km s$^{-1}$ for 4B. Converting this to
a flux density expected in the SCUBA 850\micron\ beam (assumed
bandwidth=30GHz) gives 82mJy/beam for 4A and 39mJy for 4B. This
corresponds to 1\% (4A) or 2\% (4B) of the 850\micron\ flux density
seen in our maps. This would lead to a spectral index lower by 0.01 or
0.02, not alone sufficient to explain the spectral index deficits
observed for the compact sources.  Summing the integrated intensities
for all the lines in the SCUBA 850\micron\ band gives a total value of
150 K kms$^{-1}$, still not sufficient to explain the low spectral
index of the compact sources. The outflow region may be more
susceptible to line flux contamination, as the dust continuum emission
is lower. Typical integrated line intensities measured by BSDGMA in
this region are of order 6-10K km s$^{-1}$, corresponding to a flux
level of perhaps 6 mJy/beam out of a typical continuum flux of
0.8Jy/beam in this region in the 850\micron\ map.  This is less than
1\% of the continuum flux level and so also unlikely to lead to a
significantly lower spectral index.

It should be noted that the contamination of the 850\micron\ band by CO
3-2 will be offset by the contamination of the 450\micron\ band by the CO
6-5 line. Assuming CO 6-5 is a factor of 2 stronger than the 3-2 line
(as expected if the temperature is of order 100K), the expected
spectral index dip is of order 0.005 for 4A, and the expected 
spectral index dip for the other sources is then also reduced. 

Finally, we note that there is no detectable evidence of the outflow as a
brighter region in the raw 850\micron\ map or as a less bright region
in the raw 450\micron\ map. There is an area of low emission in the outflow
region in the deconvolved 450\micron\ image (Figure~\ref{450decon}),
and of extra emission to the southwest of 4A at 850\micron.  The
450\micron\ gap could well be due to a tendency for the Richardson-Lucy
algorithm to build flux density towards the compact sources, thus clearing a
gap. It is to avoid uncertainties such as this that we have constructed the 
SI map from the raw rather than the deconvolved maps.

\subsection*{Dust opacity index}

If we can discount the preceding causes of varying $\alpha$, we are
left with the possibility that a variation in the dust opacity index
$\beta$ is responsible. This would then indicate differing dust
properties across the map.  Lower $\beta$ can be caused by larger dust
grains, indicating that grain growth has occurred in regions of lower
spectral index.  Alternatively, grain evolution from roughly spherical
to more needle-like shapes or to fluffy fractal-like structures could
have the effect of lowering $\beta$.  Chemical evolution is also
possible.  See for example Ossenkopf \& Henning (1994) or Pollack et
al (1994) for discussions of spectral index properies of evolving
grains, or Dent et al (1998) for an extensive set of observations of
young stellar objects.

The cloud in general seems to have $\beta \sim 2.3$, whilst the outflow
region has $\beta \sim 1.8$. It slightly surprising that the region of
the outflow should have such a low spectral index.  We have already
argued that line contamination alone is unlikely to explain this,
whilst drastically lower temperature in the outflow seems highly
unlikely. We suggest that dust from the region of 4A is swept up
and entrained in the outflow.

\section{Photometry and masses}
\label{phot}

Provided the emission is optically thin at some frequency $\nu$, the dust
mass of an object can be estimated by
\begin{equation}
M_{d} \approx \frac{F_{\nu}D^{2}}{\kappa_{\nu} B_{\nu}(T)}
\end{equation}
(Hildebrand, 1983),
where $F_{\nu}$ is the observed flux density, $D$ is the distance, 
$B_{\nu}(T)$ is
the Planck function for a particular temperature, and $\kappa_{\nu}$ is the dust emissivity,
given by
\begin{equation}
\kappa = 0.1 \left( \frac{250}{\lambda(\mu m)}\right) ^{\beta} 
{\rm cm^{2} g^{-1}},
\end{equation}
which we have taken from Beckwith \& Sargent (1991). Here, $\beta$ is
the dust opacity index, discussed above in Section~\ref{specin}. The
distance to IRAS4 is approximately 350pc (Herbig \& Jones, 1983).

We determined flux density levels from the deconvolved maps, by integrating
flux density within the areas indicated in Figure~\ref{regions} (the box over
4B was split into two when measuring the flux density levels of 4BI and 4BII
in the 450\micron\ map). The flux densities were calibrated in Jy/arcsec$^2$,
using the measured effective beam areas as discussed in
Section~\ref{specin}.

We estimate the masses of the compact objects using an assumed temperature
of 30K and corresponding values of $\beta$ as listed in Table~\ref{betas}.
The results for the compact components are shown in
Table~\ref{peakmass}.

\begin{table}
\begin{center}
\begin{tabular}{c|c|c|c|c|c} \hline
Wavelength & Object  & Flux density (Jy) & $\beta$ & Mass ($M_{\odot}$) \\ \hline
\vspace{0.075in}
450        & 4A,            & 139.0$\pm2.7$ &   2.1      & 11.5$^{+0.5}_{-0.5}$  \\ \vspace{0.075in}
           & 4BI,           & 62.5$\pm2.2$  &   2.3      &  5.8$^{+0.3}_{-0.3}$  \\ \vspace{0.075in}
           & 4BII,          & 16.4$\pm0.6$  &   2.3      &  1.5$^{+0.1}_{-0.07}$   \\ \vspace{0.075in}
           & 4C,            &  21.2$\pm3.8$ &   2.9      &  2.8$^{+0.4}_{-0.4}$    \\ \vspace{0.075in}
850        & 4A,            & 12.9$\pm 0.04$&   2.1  & 10.9$^{+1.1}_{-1.0}$ \\ \vspace{0.075in}
           & 4B,            & 6.4$\pm 0.05$&    2.3  &  6.9$^{+0.7}_{-0.7}$ \\ \vspace{0.075in}
           & 4C,            & 1.3$\pm0.04$ &    2.9  &  2.9$^{+0.3}_{-0.3}$  \\ \vspace{0.075in}

\end{tabular}
\caption{\label{peakmass} Dust masses derived from the photometry for
the compact sources. Total flux densities are measured by integrating
the flux density within an aperture - see Figure~\ref{regions} for the
apertures used in the case of 450\micron. The temperature is assumed
to be 30K in each case.  Source 4B has been considered as two separate
sources at 450\micron, but not at 850\micron. Quoted errors for the
flux densities are the same percentage error for each source as the
flux density error in Table~\ref{rawflux}, and so contain no estimate
of calibration uncertainties, uncertainties due to the choice of
aperture, etc. Errors for the masses are a combination of the
photometric errors, and an error of $\pm$0.08 in the dust opacity
index $\beta$, as measured from the cloud region between the compact
sources. Note that although the real error in the absolute value of
$\beta$ is larger than this, it is not independent of the assumed
temperature. The spectral index uncertainty usually dominates,
particularly at 850\micron. The variation due to different assumptions
of temperature is considerable. For example, adopting an alternative
temperature of 50K would lead to a reduction in the derived masses by
a factor of approximately 2. }
\end{center}
\end{table}

\subsection{Extended emission}
\label{extended}

Estimating the mass of the extended emission is problematic.
For a start, the cloud extends beyond the edges of the 
jiggle maps, so a direct measurement of the total cloud emission
and mass is impossible. Another problem arises from the danger 
that parts of the map may chop onto other parts. The chop direction 
is approximately in the $\pm$ RA direction, and the throw is 120''.
Thus the central regions should chop onto regions away from the
extended ridge. The areas at upper right and lower left may well be 
affected by chopping onto significant emission, however.

\begin{table}
\begin{center}
\begin{tabular}{c|c|c|c|c} \hline
$\lambda$ &  Flux density & T & $\beta$ & Mass (M$_{\odot})$ \\ \hline
450       & 47.5$\pm5.2$  & 50  & 2.3      &  2.1$\pm0.3$     \\
          &               & 100 & 2.1      &  0.78$\pm0.09$   \\ \hline
850      &   7.73$\pm0.08$& 50  & 2.3      &   4.4$\pm0.5$ \\ 
         &                & 100 & 2.1      &  1.6$\pm0.2$ \\ 
\end{tabular}
\caption{\label{extmass} Cloud masses for different assumptions of 
temperature and corresponding $\beta$ at 450\micron\ and 850\micron.
Photometric errors are calculated based on the measured background RMS
deviation from the mean, and include contributions from the uncertainties
in the subtracted core flux densities. Errors for the masses
also include the effect of a $\pm 0.08$ random error in the opacity index $\beta$. }
\end{center}
\end{table}

We can make a measurement of the extended flux density in the maps, in
order to produce a lower limit to the total mass of the surrounding
cloud. This was done by summing the emission over the central part of
the map (shown in Figure~\ref{regions}) and subtracting the flux
already determined for the compact sources.  Table~\ref{extmass} shows
the masses calculated from the extended emission at 450\micron\ and
850\micron\ for various assumed parameters. These masses are
considerable, representing about half the mass of the system.  The
850\micron\ mass estimates are about a factor of 2 greater than those
at 450\micron.

\subsection{Mass ratios}
\label{ratios}

Calculating mass ratios at 450\micron\ and 850\micron\ from the data in
Table~\ref{peakmass} and averaging, we obtain;
\begin{equation}
\frac{4BI + 4BII}{4A} = 0.63,
\end{equation}
and
\begin{equation}
\frac{4C}{4A + 4BI + 4BII} = 0.15.
\end{equation}
The ratio for the 4BI - 4BII system, measured from the 450\micron\
 map only, is
\begin{equation}
\frac{4BII}{4BI} = 0.26.
\end{equation}

\section{Discussion}

\subsection*{Morphology of IRAS4}

The elongation of 4A could be explained as a circular disk inclined at
an angle of approximately 53$^{o}$ to the plane of the sky.  
The elongation of 4B in the raw map appears to be EW. However, upon
deconvolving the beam from the map we discovered that 
this elongation is due to 4B being itself a binary system. 
The primary component of 4B in the deconvolved 450\micron\ map
appears to have a slight elongation along the 
same axis as 4A. 

The amount of diffuse material in which the system is embedded is
difficult to determine with any accuracy, since the map does not
encompass the entire extended ridge, the flux density levels may be affected
by the chop throw sampling some of the extended emission, and at
450\micron\ the error beam may spread flux density from the compact sources
into the surrounding regions. 

\subsection*{Stability of the system}

The separation of the 4B double is approximately 12'' on the sky. If
we assume that the triple system is coplanar, so that the 4BI-BII
axis has the same inclination to the plane of the sky as the 4A disk,
the true BI-BII separation would then be 16'' (5600 AU for
a distance of 350pc). The 4A-4B separation is approximately 30''.  The
long axis of the 4A disk extends some 8'' from the centre in the
deconvolved images (we measure the disk extent as the distance at
which the flux density reaches 5\% of its peak value). Thus the
presence of the 4A disk is not incompatible with the interpretation
that 4A, 4BI and 4BII have coplanar orbits.

\begin{figure}
\psfig{{figure=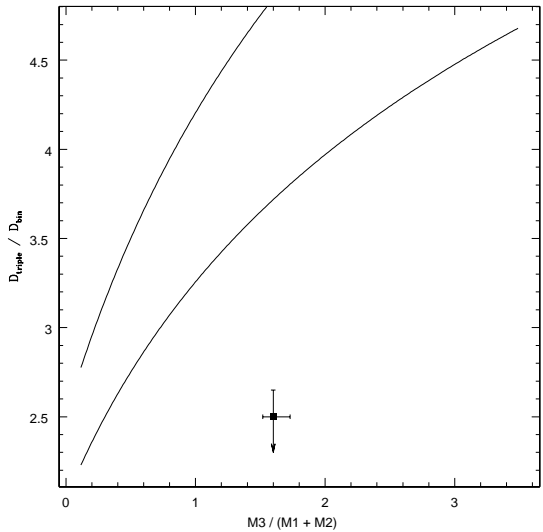,width=3.truein,height=3.truein}}
\caption{\label{stabfig} Harrington's stability test as applied to
IRAS4.  The x-axis is the mass ratio $M_3 / (M_1 + M_2)$. The y-axis
is the ratio of periastron distance to binary separation. The two bold
lines represent the corotating and counterrotating stability
limits. The lower of the two represents the more stable
counterrotating case. Systems above these lines will be stable.  The
position of the 4A-4BI-4BII system is marked as a point. The error 
in the mass ratio is calculated from the errors in Table~\ref{peakmass}.
The uncertainty in the ordinate has been estimated as follows. The 
misalignment between component BI and the axis of elongation of 4A
was projected back into the plane of the supposed 4A disk. This then gives
the possible discrepency between the 4A-4BI distance measured in the plane of 
the sky and the inferred 4A-4BI distance in the plane of the disk. 
The upper error bar shows the possible increase in stability 
due to this observed misalignment. Since the quantities $D_{triple}$ and 
$D_{bin}$ have been chosen to maximise stability, the y-position
of the IRAS4 system is an upper limit.}
\end{figure}

We can also make an assessment of how stable the triple 4A-4BI-4BII
system is. We first assume that the system is coplanar and the orbits circular.
A criterion for stability in a hierarchical triple was
developed by Harrington (1977), 
\begin{equation}
\frac{D_{triple}}{D_{binary}} > K \left\{ 1 + A 
\ln \left[\frac{2}{3} \left( 1 + \frac{M_3}{M_1 + M_2} 
\right) \right] \right\}.
\end{equation}
Here, $D_{triple}$ is the periastron distance of the stand-alone star,
and $D_{binary}$ the semi-major axis of the binary orbit. $M_3$ is the
mass of the single component, $M_1$ and $M_2$ the masses of the binary
components. The constants have values $K=3.5$ and $A=0.7$ for a
corevolving system, or $K=2.75$, $A=0.64$ for a counterrotating
system. For our case, the mass ratio is $1/0.63$
(Section~\ref{ratios}).  The semimajor axis of the BI-BII system
cannot be smaller than the observed separation on the sky of 12''. The
true 4A-4BI separation of course depends on the angle of the
binary-single star axis.  We have already argued (as have previous
authors) that the 4A elongation is an inclined disk.  If we assume
that the 4A-4BI orbit is coplanar with the 4A disk (expected if
fragmentation formation models apply), then the true 4A-4BI distance
would be 30'', which would also be the largest possible
periastron distance. Taking these figures we find that the system fails
Harrington's stability test easily, even if we consider the more
stable counterrotating case.  See Figure~\ref{stabfig}.

It is possible to envisage accretion of material stabilising
initially-unstable triple systems by modifying the separation ratios
of the components (Smith et al, 1997). There certainly seems to be a
substantial reservoir of available material in the IRAS4 envelope (see
Table~\ref{extmass}). In the case of a low-mass binary orbiting a
higher-mass single star, as here, this mechanism is unlikely to result
in stability.  Low specific angular momentum material will tend to be
accreted onto IRAS4A, causing the single-binary separation to decrease
and destabilizing the system. High specific angular momentum will tend
to accrete onto the binary (4BI-4BII), widening it by adding angular
momentum and again destabilizing the system. Furthermore, the
4A-4BI-4BII system fails the Harrington stability test by a
comfortable margin. Even if the available envelope mass were accreted
exclusively onto 4BI and BII, moving the mass ratio towards 1, there is
not enough mass to entirely stabilize the system. We conclude that
IRAS4 seems to be an unstable multiple which will be disrupted within
a few orbits.

\subsection*{Star forming history of IRAS4}

Could the elongation of the main cloud be explained as a foreshortened
disk, in a similar way to the elongation of 4A? The young age (approx
10$^5$ years deduced from the embedded nature), appears to preclude
this interpretation. For a circumstellar disk to form requires several
of the disks dynamical time and a disk of this size would have a
period in excess of 4$\times$10$^5$ years. This implies that the extended
structure must be a remnant of the cloud's initial conditions and not
due to the subsequent collapse dynamics.

Based on their polarimetric observations, Minchin et al suggested that
the extended emission is an inclined ``pseudo-disk'', of the sort
envisaged by for example Galli \& Shu (1993). In this case, the cloud
material is magnetically supported in one direction, but free to
collapse in the other. A massive, non-centrifugally supported disk can
therefore form in a shorter time than would be expected on dynamical
grounds. The ambipolar diffusion timescale would have to be longer
than the freefall time of the cloud, and the magnetic field would be
an impediment to fragmentation of the central region, because 
it would tend to enforce solid body rotation. For this reason, we do not
favour the ``pseudo-disk'' interpretation.

Simulations indicate that a prolate cloud, with an end-over-end
rotation, should undergo fragmentation and form a central binary
(Bonnell et al, 1992, Bonnell \& Bastian 1992). The additional
multiplicity of the system (4BI-4BII) is then explainable as being due
to an internal disk fragmentation (Bonnell 1994) that occurs in the
individual components formed in the prolate cloud fragmentation
(Bonnell et al 1992).  This interpretation still leaves us with some
unexplained details. Most importantly, the third component to the
north east. There are several possible explanations for this
component. Firstly, it could be part of the IRAS 4 system, but lie
outside of the prolate cloud, in front of the main system and well
above the plane of the IRAS 4A disk and inferred 4A-4B orbit.  This
poses substantial problems for fragmentation models where collapse
occurs preferentially in one direction, reducing the dimension of the
cloud and making it unstable to fragmentation (Bonnell~1999).
Secondly, it could lie in the same plane as the 4A disk, which would
place it well outside the prolate cloud as it is seen in the maps, at
a distance of 15,400 AU (44'' at 350pc).

If IRAS 4C is part of the IRAS 4 system, then the most
promising explanation is that an independent condensation (see eg,
Pringle~1991) was present near the prolate cloud when collapse occured.

There exists the possibility that IRAS 4C is not a member of the IRAS
4 system but just a chance projection. Its separation and the presence
of other sources nearby lend credence to this possibility, as does
it's spectral index being higher than the surrounding cloud.  A
determination of the velocities of the various components could shed
further light on the possible relationship of 4C with the central
system.


\section{Summary}

We have examined high signal to noise maps of the protostellar
multiple system NGC1333/IRAS4. We have estimated the masses of the
principle components, and discussed the probable configuration,
possible history, and likely future of the system on dynamical
grounds.  In the light of fragmentation models of binary formation, we
argued that the elongated geometry of the main cloud is probably due
to a prolate rather than disk-like geometry, because
magnetically-supported flattened structures are not expected to
undergo fragmentation to form a multiple system such as IRAS4. The
third major component, 4C, poses problems for any almost any model as
it lies either too far from the central system, or out of the plane of
the main binary orbit. We suggest that 4C may be an independent
condensation in the cloud. By comparing the masses and separations of
the central triple system, assuming coplanarity, we showed that it is
clearly not stable. Thus IRAS4 is an example of a multiple system
which has formed from fragmentation of a tumbling prolate cloud, but
which is will in future disintegrate to leave fewer, lower order
multiple or single systems.

We constructed a map of spectral index for the system. Despite the
various problems associated with the measurement of this quantity, we
concluded that there is marginally significant evidence that 4A
has a low spectral index compared to the cloud. There is also strong
evidence that the region associated with the 4A outflow has a low
spectral index. Various possible causes of this were
discussed. There is no evidence suggesting that the emission is
optically thick at either wavelength. The apparently low spectral
index of 4A was dependent on temperature assumptions, but the outflow
region should not be systematically cooler than the surrounding cloud,
so this is unlikely to provide an explanation. We considered the
possibility that molecular line emission could manifest itself as a
region of low spectral index by contaminating the 850\micron\ flux
density. Based on the measured line strengths from BSDGMA, this
explanation seemed unlikely to be the sole cause, although it could
constitute a partial cause. It was therefore
concluded that the low spectral index of 4A was most likely an
indication of dust grain growth in the dense circumstellar disks. This
grain growth then appears to be more advanced in 4A than in 4B or 4C.

We have seen an area of low spectral index running through IRAS4A, in
the position of the molecular outflow seen by BSDGMA.  This ridge is
seen on both sides of IRAS4A, and is clearly distinct from IRAS 4C to
the Northeast. We argued in Section~\ref{lines} that contamination
from molecular line flux is probably not sufficient to explain this.
We suggest the possibility that dust from the vicinity of IRAS4A is
being swept up and entrained in the outflow.  This suggests models in
which the driving mechanism for the outflow is a jet embedded in the
protostellar core itself (see e.g.  Masson \& Chernin 1993; Chernin et
al 1994; Raga \& Cabrit, 1993). If this explanation is correct, the
main issues for outflow theories would seem to be that the driving
mechanism should not be so violent that dust grains are destroyed in
large numbers, and that significant amounts of dust material should
occupy the outflow cavity after the main jet working surfaces have
passed through.

\section{Acknowledgements}

The JCMT is operated by the Joint Astronomy Centre, on behalf of the UK
Particle Physics and Astronomy Research Council, the Netherlands
Organization for Pure Research, and the National Research Council of
Canada. The authors thank James Deane 
and Jane Greaves for helpful discussions and advice.

\label{lastpage}

\end{document}